\documentclass[a4paper,fleqn,usenatbib]{mnras}

\usepackage{mathptmx}
\usepackage{txfonts}

\usepackage[T1]{fontenc}
\usepackage{ae,aecompl}

\usepackage{graphicx}	% Including figure files
\usepackage[]{units}
\usepackage{longtable,lscape}
\title[The young open cluster NGC 7067]{The young open cluster NGC 7067 using Str\"omgren photometry}

\author[M. Mongui\'o et al.]{
M. Mongui\'{o}$^{1,2}$\thanks{E-mail: m.monguio@herts.ac.uk},
I. Negueruela$^{1}$,
A. Marco$^{1,12}$,
C. Gonz\'alez-Fern\'andez$^{3}$,\newauthor
J. Alonso-Santiago$^{1}$, 
M. T. Costado$^{4}$,
L. Casamiquela$^{5}$,
M. L\'opez-Corredoira$^{6,7}$,\newauthor
J. Molg\'o$^{8,6}$,
F. Vilardell$^{9}$,
E. J. Alfaro$^{4}$,
T. Antoja$^{10}$,
F. Figueras$^{5}$,
M. Garcia$^{11}$,
\newauthor
C. Jordi$^{5}$,
and M. Romero-G\'omez$^{5}$
\\
$^{1}$Dpto. de F\'{\i}sica, Ingenier\'{\i}a de Sistemas y Teor\'{\i}a de la Se\~{n}al. Escuela Polit\'{e}cnica Superior, Universidad de Alicante.\\
$^{}$\ Campus de Sant Vicent del Raspeig, s/n. 03690 Sant Vicente del Raspeig (Alicante). Spain. http://dfests.ua.es. \\
$^{2}$School of Physics, Astronomy \& Mathematics, University of Hertfordshire, College Lane, Hatfield, Hertfordshire, AL10 9AB, U.K.\\
$^{3}$Institute of Astronomy, University of Cambridge, Madingley Road, Cambridge, CB3 0HA, UK.\\
$^{4}$Instituto de Astrof\'isica de Andaluc\'ia-CSIC, Apdo. 3004, 18080 Granada, Spain. \\ %3 granada
$^{5}$Departament de F\'isica Qu\`antica i astrof\'isica and IEEC-ICC-UB,  Universitat de Barcelona, Mart\'i i Franqu\`es, 1, 08028 Barcelona.\\ %2 ub
$^{6}$Instituto de Astrof\'isica de Canarias, 38205 La Laguna, Tenerife, Spain.\\ %4 iac
$^{7}$Departamento de Astrof\'isica, Universidad de La Laguna, 38206 La Laguna, Tenerife, Spain.\\ %5 laguna
$^{8}$GRANTECAN, Cuesta de San Jos\'e s/n, E-38712 , Bre\~na Baja, La Palma, Spain. \\	%6 gtc
$^{9}$Institut d'Estudis Espacials de Catalunya, Edifici Nexus, c/ Gran Capit\`{a}, 2-4, desp. 201, 08034, Barcelona.\\ %7 IEC
$^{10}$Research and Scientific Support Office, European Space Agency (ESA-ESTEC), PO Box 299, NL-2200 AG Noordwijk, the Netherlands. \\ %8 esa
$^{11}$Centro de Astrobiolog\'ia, CSIC-INTA. Ctra. Torrej\'on a Ajalvir km.4, E-28850 Torrej\'on de Ardoz, Madrid, Spain. \\ 
$^{12}$ Department of Astronomy, University of Florida, 211 Bryant Space Science Center, Gainesville, FL 32611. %Florida
}

\date{Accepted 2016 December 14. Received 2016 December 13; in original form 2016 October 14}

\pubyear{2016}

\begin{document}
\label{firstpage}
\pagerange{\pageref{firstpage}--\pageref{lastpage}}
\maketitle

\begin{abstract}
NGC 7067 is a young open cluster located in the direction between the first and the second Galactic quadrants and close to the Perseus spiral arm. 
This makes it useful for studies of the nature of the Milky Way spiral arms.
Str\"omgren photometry taken with the Wide Field Camera at the Isaac Newton Telescope allowed us to compute individual physical parameters for the observed stars and hence to derive 
cluster's physical parameters.
Spectra 
from the 1.93-m telescope at the Observatoire de Haute-Provence helped to check and improve the results.
We obtained photometry for 1233 stars, individual physical parameters for 515 and spectra for 9 of them. The 139 selected cluster members lead to a cluster distance of 4.4$\pm0.4$\,kpc, 
with an age below $\log_{10} (t(\unit{yr}))$=$7.3$ and a present Mass of 1260\,$\pm$\,160\,$\unit{M}_{\sun}$. 
The morphology of the data reveals that the centre of the cluster is at $(\alpha,\delta)$=$(21\colon24\colon13.69,+48\colon00\colon39.2)$ J2000, with a radius of 6$\farcm$1. 
Str\"{o}mgren and spectroscopic data allowed us to improve the previous parameters available for the cluster in the literature.%, giving more accurate results over a larger sky area. 

\end{abstract}

\begin{keywords}
Methods: observational --
Open clusters and associations: individual: NGC 7067 --
 Hertzsprung-Russell and C-M diagrams
\end{keywords}

%%%%%%%%%%%%%%%%%%%%%%%%%%%%%%%%%%%%%%%%%%%%%%%%%%

%%%%%%%%%%%%%%%%% BODY OF PAPER %%%%%%%%%%%%%%%%%%

\section{Introduction}\label{sec:intro}
Open clusters are a good framework to study stellar and Galactic evolution. The assumption that all the stars belonging to a cluster were born at the same time provides us a snapshot of the stellar evolutionary process
for different stellar masses, which is key to analyse stellar evolution. 
Several authors have made use of them to analyse the structure of the Milky Way, and in particular, of its spiral pattern. 
Already in the 60's \citet{1961LowOB...5..133J} studied their distribution in the Galactic disc. 
Later on, \citet{1982ApJS...49..425J} discussed their vertical distribution and the spiral pattern of the Milky Way they traced. % in the Milky Way.  
\citet{1989A&A...209...51K} made a deeper analysis  of the clusters in the second Galactic quadrant finding a structure probably related with the Perseus arm.
\citet{2006MNRAS.368L..77M} and \citet{2008ApJ...672..930V} are other examples of authors who used the clusters  to study the spiral arms location.
 \cite{2013MNRAS.432.3349C} used up to 27 open clusters in the anticentre region to fit the loci for the Perseus arm, and also to
confirm the presence of the outer arm along the second and third galactic quadrants.
More recently, \citet{2015MNRAS.449.2336J} used them to  trace the  pattern speed of spiral structure.

%PREVIOUS SOBRE EL CUMUL
NGC 7067 is a young open cluster located between the first and the second Galactic quadrants $(\ell,b)$=(91.2, -1.7)$\degr$, and very close to the Perseus spiral arm.
\citet{1961PUSNO..17..343H} presented $UBV$ photoelectric measurements for 29 stars in the field. 
Using these data, \citet{1961LowOB...5..133J} established a distance of 2900\,pc to the cluster, using colour-magnitude and colour-colour diagrams. 
But the same authors claimed that this distance was the most poorly determined of their list of 106 clusters. 
\citet{1963ZA.....57..117B}, only from the colour-magnitude diagram of these same data, derived 3950\,pc. Afterwards, \citet{1965MmSAI..36..283B} tried to improve the distance determination to the cluster 
adding the two colour-colour magnitude diagram information of $\sim$60 stars using photographic plates, and obtained 4500\,pc. However, he found some discrepancies between the $V-(U-B)$ and $V-(B-V)$ diagrams.
 \citet{1965ApJS...12..215H} used calibrations that provide absolute magnitudes from H$\gamma$ equivalent widths for only three early-type stars in NGC 7067 obtaining a distance of 2290\,pc. They also
used spectra for  these three stars to obtain a spectroscopic distance of 2190\,pc.
\citet{1968ArA.....5....1L} provided an age for the cluster of $\log_{10} (t(\unit{yr}))$=7.1, based on the absolute magnitude distribution of the brightest 8 stars in the cluster.
The mentioned studies are based on few stars. \citet{1973A&AS....9..261H} increased the statistics using $UBV$ photometry for 111 stars from photographic plates, calibrated using \citet{1961PUSNO..17..343H} photometry.
From the two colour-magnitude diagrams he obtained a distance of 3963\,pc, although when he used only the combined data of the stars in common with \citet{1961PUSNO..17..343H}, the 
value obtained was 4406\,pc. 
Since no turn off point was found in the data, the cluster age was estimated to be smaller than 10\,Myr.
% No supergiant stars were found in the data, so no clear turn off point can be seen in the diagrams, but the presence of a B0.5V star was used to claim an
% upper limit for the age of 10\,Myr.

\citet{1983A&A...121..237B} estimated the absolute mass of the cluster: $M_{\unit{C}}$=550$\unit{M}_{\odot}$. They set a zero point using three clusters for the relative masses obtained by 
\citet{1978stfo.book.....R}, who used $UBV$ data and assumed
a stellar mass distribution function \citep{1955ApJ...121..161S} and a universal mass-luminosity relation for the clusters.
\citet{2004MNRAS.349.1481Y}  were the first to cover the cluster with CCD observations, extending the photometry to redder bands using $UBVRI$ filters  and obtaining photometry for almost 2000 stars in the field. 
The colour-colour diagrams, including available 2MASS data, allowed them to obtain several physical parameters like  a distance of 3.6$\pm$0.2\,kpc, and
a mean colour excess of $E(B-V)$=0.75$\pm$0.05\,mag. By fitting ZAMS at different metallicity for these data, the best fit was obtained for $Z\sim0.02$.
The isochrone fitting in the colour-magnitude diagram provided an age of  $\log_{10} (t(\unit{yr}))$=8. Although this last study used more stars and a wider range of filters, the obtained
 age is considerably older than previous studies, and it is in fact not compatible with the presence of early-B stars.
 All previous studies have used broad band photometry (and spectra for just three stars) with distances ranging from 2.2 to 4.5\,kpc.

In the current paper we present new intermediate band photometry for 1233 stars in the field and spectroscopic data for 9 of them. The $uvby\beta$ Str\"omgren  CCD photometry \citep{1966ARA&A...4..433S} 
provides more accurate information than usual Johnson $UBV$ photometry, since the bands are narrower. This system is the most
adequate for the study of early-type stars, since it has been purposely designed to provide accurate measurements of their intrinsic properties.
Besides the colour-magnitude plots, these data allow us to compute physical parameters
for individual stars \citep{1978AJ.....83...48C,2014A&A...568A.119M}. 
In addition, the spectra available for few stars in the cluster allow us to check the consistency of the results.

In Sect.\,\ref{sec:Obs} we describe the photometric and spectroscopic observations as well as the data available. In Sect.\,\ref{sec:Res} we analyse these data from different perspectives. 
In Sect.\,\ref{sec:spectra} we discuss the spectral type of the stars having spectroscopic observations.
In Sect.\,\ref{sec:photdiag} we analyze different colour-magnitude and colour-colour diagrams. 
The selection of cluster members is developed in Sect.\,\ref{sec:membership}, where the parameters of the clusters are discussed.
Individual physical parameters  for hot stars are computed in Sect.\,\ref{sec:FP} while in Sect.\,\ref{sec:other} we use IPHAS and 2MASS photometric data to detect outliers and emission line stars.
We analyse the cluster density, its age and Mass in Sect.\,\ref{sec:density} and \ref{sec:Sectage}, respectively. 
In Sect.\,\ref{sec:disc} we discuss the relation between the cluster and the Perseus spiral arm, and the importance that the analysis of clusters like this may have in the study of the nature of the spiral arms of the Galaxy.
Finally, in Sect.\,\ref{sec:conclusions} we summarize the results.

\section{Observations and data}\label{sec:Obs}

\subsection{Str\"omgren photometry}

\begin{figure}\centering
 \resizebox{\hsize}{!}{\includegraphics{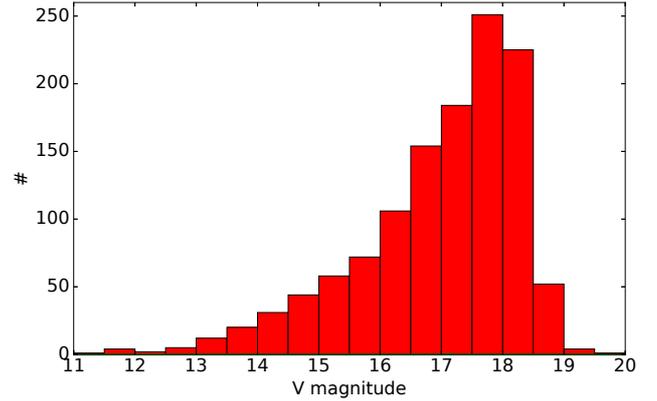}}
 \caption{ $V$ magnitude histogram for all the observed stars.}\label{fig:Vhist}
\end{figure}

The  Wide Field Camera (WFC) at the Isaac Newton Telescope (INT, 2.5m) --located at El Roque de los Muchachos in the Canary
Islands-- was used to obtain Str\"omgren $uvby\beta$ CCD photometry for this cluster.
 The WFC is a 4 chip mosaic of thinned AR coated EEV 4K$\times$2K devices
with pixels size of 0$\farcs$333 and an edge to edge limit of the mosaic of 34$\farcm$2. For the study of this cluster, only data from the central chip has been analysed.
The six filters used were Str\"{o}mgren $u$, $v$, $b$, $y$, $H\beta _w$, $H\beta_n$. No pixel binning and fast read-out mode were used for the
observations.
The observations were acquired during the nights of 2014 July 6 to 8.
Different exposure times for each filter were used, in order to cover the full range of visual magnitudes needed, avoiding saturated stars and 
reaching a limiting magnitude around 18\,mag (see Fig.\,\ref{fig:Vhist}). The used exposure times can be found in Table \ref{tab:exptime}.
%calibration
The central chip (number 4) was used and calibrated 
through the standard clusters NGC 6910 \citep{1977AJ.....82..606C}, NGC 884 and NGC 869 
\citep{1970AJ.....75..822C,1955ApJ...122..429J}. See the photometry for the standard individual stars in \citet{2013A&A...552A..92M}.
%prereduction
The images were reduced using several \textsc{iraf}\footnote{\textsc{iraf} is distributed by the National Optical Astronomy Observatories,
    which are operated by the Association of Universities for Research     in Astronomy, Inc., under cooperative agreement with the National
    Science Foundation \citep{1986SPIE..627..733T}.} tasks. Firstly, the bias
derived from the overscan areas was subtracted and the bad pixels were replaced  by linear interpolation using the nearest good pixels through the \textsc{fixpix} task.
 Flatfielding was applied using the sky flats obtained during the whole observing run. 
All the stars available in
the images were located using the \textsc{daofind} routine.
The instrumental photometry was obtained through twelve different aperture radii that provided
twelve different magnitudes for each star. The \textsc{mkapfile} routine uses the \textsc{daogrow} algorithm to obtain the aperture corrections and
the fitted radius. The final instrumental magnitudes were computed from the integration of the obtained curve of growth \citep[see][for details]{1990PASP..102..932S}.
We corrected from atmosphere extinction using the median extinction coefficients for the La Palma observatory.
The transformation to the standard system was developed using %the observed standard clusters and 
the transformation equations described in \citet{2013A&A...549A..78M}.
We computed the transformation coefficients using the data from the whole run from July 3 to July 8.
The obtained coefficients and the different transformation equations are:
\begin{eqnarray}\nonumber
y'-V_{\unit{cat}}&=& (-23.89\pm 0.01) +(-0.08\pm 0.02) \cdot (b-y)_{\unit{cat}} \\\nonumber
(b-y)'&=&(-0.197\pm  0.008) +  (0.93\pm 0.02)(b-y)_{\unit{cat}} \\\nonumber
c'_1&=&(0.17\pm0.01)+(-0.16\pm0.01)\cdot(b-y)_{\unit{cat}} +\\\nonumber
&&( 0.95\pm0.01)\cdot c_{1\unit{cat}}\\\nonumber
(v-b)'&=&(0.361\pm0.007)+( -0.015\pm0.008)\cdot(b-y)_{\unit{cat}} + \\\nonumber
&&(1.06\pm0.01)\cdot(v-b)_{\unit{cat}}+ (0.012\pm 0.018)\cdot c_{1\unit{cat}}\\
% m_{1\unit{st}}=(v-b)_{\unit{st}}-(b-y)_{\unit{st}}$
\beta'&=& (2.30\pm0.03)+ (0.94\pm0.02)\cdot (\beta_{\unit{cat}}-2.8),
\end{eqnarray}
where the prime indicates instrumental extinction-corrected magnitudes and the subscript ${\unit{cat}}$ indicates the standard catalogued values. 
Then, for the scientific targets, the standard values are obtained inverting the previous equations, and taking into account that $m_{1}=(v-b)-(b-y)$.

\begin{figure*}\centering
 \resizebox{\hsize}{!}{\includegraphics{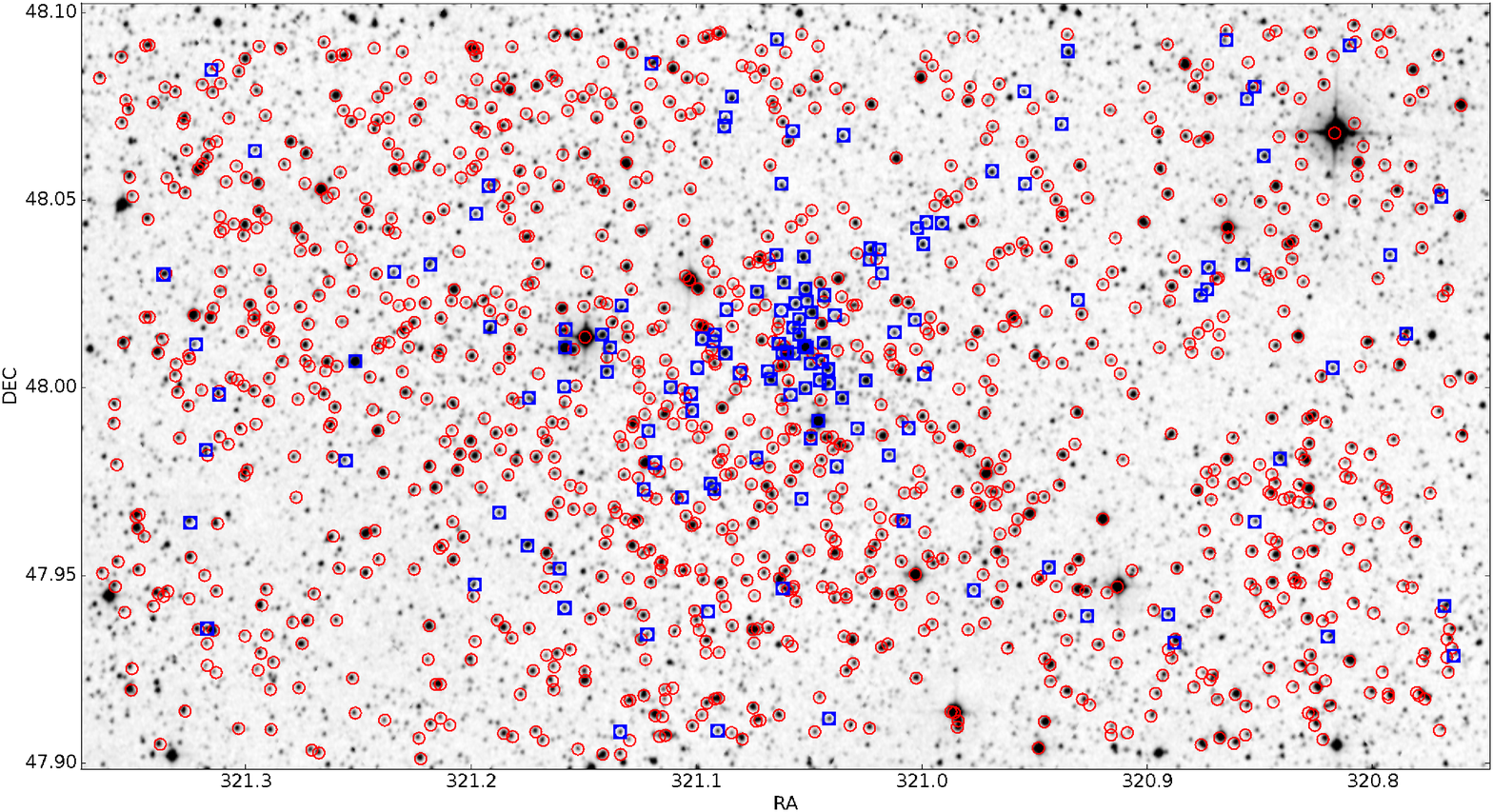}}
 \caption{Finding chart of the 1233 stars with available Str\"omgren photometry (red circles). Blue squares are the stars classified as likely members in this study (see Sect.\,\ref{sec:membership}). 
The underlying image is taken from DSS2-F North survey.}
\label{fig:FC1}
\end{figure*}

We obtained $V, (b-y), m_1, c_1$, and $\beta$ for 1233 stars in the field (see their sky distribution in Fig.\,\ref{fig:FC1}). These stars were observed up to 10
 times in each filter, with different exposure times.
The final photometry for each target was obtained through a weighted mean, taking into account that:
1) the weight applied was $w_i$=$1/\sigma_i^2$  where $\sigma_i$ is the 
individual error for each index, computed with full propagation errors (and so taking into account the individual exposure times);
2) outliers were rejected using a 3$\sigma$ rejection process, obtaining a final number of measurements different for each index, and
3)  the error of the mean was computed for each index. 
The final right ascension and declination coordinates for each target were  computed 
also using mean values from all the measurements. 
The positions in the J2000 coordinate system were determined using \textsc{wcs} (a routine from the WCSTools package\footnote{http://tdc-www.harvard.edu/software/wcstools/ }) 
and a fifth-order polynomial taking USNO-B1 \citep{2003AJ....125..984M} as reference catalog.
The final values are available in Table \ref{tab:TablePhot}. A random number is assigned as identifier for each star.

\begin{figure}\centering
 \resizebox{\hsize}{!}{\includegraphics{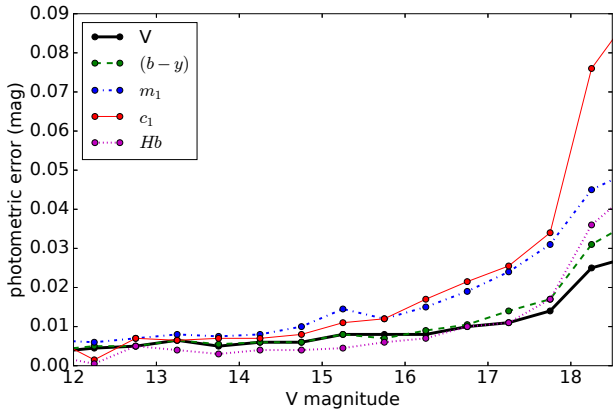}}
 \caption{Internal errors in photometric indexes as a function of $V$ magnitude. Median values have been computed for 0.5\,mag bins.}\label{fig:errPhot}
\end{figure}

Figure \ref{fig:Vhist} shows the $V$ magnitude distribution of the observed stars.
We computed the limiting magnitude as the 
mean of the magnitudes at  the  peak star  counts in  the  magnitude histogram and
its  two  adjacent  bins,  before and  after  the  peak,  weighted  by the  number of  stars  in  each  bin. We obtained $V_{\unit{lim}}$=17.8\,mag.
\citet{2013A&A...549A..78M}  estimated  that  the  limiting $V$ magnitude computed with this simple algorithm provides
the $\sim$90 per cent completeness limit.
We show in Fig.\,\ref{fig:errPhot} the internal errors as a function of magnitude for the different photometric indexes. 
Individual errors are computed as the error of the mean for all the available measurements when two or more measurements are available. When all the measurements are taken consecutively (during the same night) 
those errors could be slightly underestimated.
When only one measurement is available, then the error provided is computed by error propagation from the instrumental measurements and the transformation equations.
In the plot we show the median value for each 0.5\,mag bin, obtaining values below 0.03\,mag for stars up to $V$=18\,mag, being slightly larger for the $c_1$ index 
(since it contains the $u$ magnitude, that always gives larger errors).

\begin{table}
 \caption{Central wavelengths and bandwidths of the filters. Last column shows the exposure times.}\label{tab:exptime}
\centering\begin{tabular}{c|cc|c}
filter& $\lambda_0$(\AA{})&$\Delta \lambda$(\AA{})&Exp.time (s)\\\hline
$u$	&3480&330	&30, 600, 1200, 1800\\
$v$	&4110&150	&12, 150, 600 \\
$b$	&4695&210	&6, 60, 350\\
$y$	&5500&240	&5, 30, 100, 300\\
$H\beta _w$&4861&170	&5, 50, 100, 300\\
$H\beta _n$&4861&30	&30, 600, 1200, 1500\\
 \end{tabular}
 \end{table}

\subsection{Spectroscopy}

Spectra of nine stars in the cluster were obtained in October 2001, using the 1.93\,m telescope at the Observatoire de Haute Provence
(OHP), France. 
Those were selected as the brightest stars whose positions in published photometric diagrams suggested they could be blue stars.
The telescope was equipped with the long-slit spectrograph {\em Carelec} \citep{1990A&A...228..546L} and a 1024$\times$2048 pixel EEV CCD.
Some of the brightest stars of the cluster were observed on the nights of October 23rd and 24th using the 600 ln/mm grating. 
This configuration gives a nominal dispersion of $\sim0.9$\AA/pixel over the 3750--5570\AA\ range. 
A few more stars were observed on October 25th with the 300 ln/mm grating, giving a nominal dispersion of $\sim1.8$\AA/pixel over the 3600--6900\AA\ range. 
Given the impossibility to assign a spectral type to the bright star LS\,III~$+47$\,36 (Star $\sharp$2 in the \textsc{webda} notation, ID$_{{\unit {WD}}}$ from now on), we re-observed it on two occasions with the INT equipped with the  
Intermediate Dispersion Spectrograph (IDS) and the 235-mm camera. 
The first observation was on 2002 July 25th, using the R1200Y grating and a Tek\#4 camera, which gives a dispersion of $\approx0.8$\AA/pixel. 
The second observation took place on 2003 July 5th. On this occasion, the R900V grating and EEV\#13 camera were used, resulting in a nominal dispersion of $\approx0.65$\AA/pixel. 
Exposure times and SNR for those spectra can be found in Table \ref{tab:specSNR}.

\begin{table}
 \caption{Observing date, exposure time and signal to noise ratio for the available spectra. See more details in the text}\label{tab:specSNR}
\centering\begin{tabular}{lcccc}
Name&ID$_{{\unit {WD}}}$  & Date & Exp.time (s)&SNR\\ \hline
LS\,III~$+47$\,39&4    &24 Oct 2001&1500&115\\
                 &9    &23 Oct 2001&1200&75\\
LS\,III~$+47$\,38&5    &23 Oct 2001&1200&110\\
                 &15   &25 Oct 2001&900&60\\
LS\,III~$+47$\,37&2012+2015 &23 Oct 2001&1200&120\\
                 &22   &25 Oct 2001&600&55\\
LS\,III~$+47$\,36&2    &25 Oct 2001&600&160\\
LS\,III~$+47$\,36&2    &25 Jul 2002&400&80\\
LS\,III~$+47$\,36&2    &5 Jul 2003&600&105\\
                 &96   &23 Oct 2001&1200&30\\
\end{tabular}
\end{table}

\begin{figure*}\centering
 \resizebox{\hsize}{!}{\includegraphics[angle=270, trim=60bp 110bp 215bp 50bp,clip]{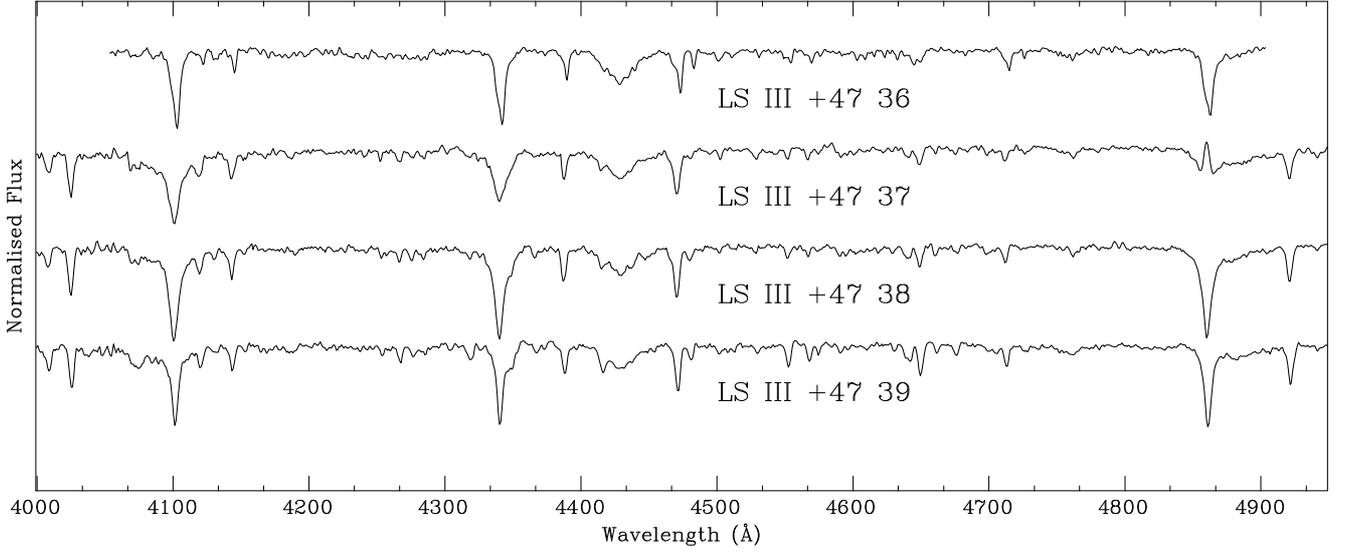}}
 \caption{Four of the available spectra. From top to bottom their WEBDA identifier is ID$_{{\unit {WD}}}$=$\sharp$2, $\sharp$2012+$\sharp$2015, $\sharp$5, and $\sharp$4.}\label{fig:spec}
\end{figure*}

\begin{figure}\centering
 \resizebox{\hsize}{!}{\includegraphics{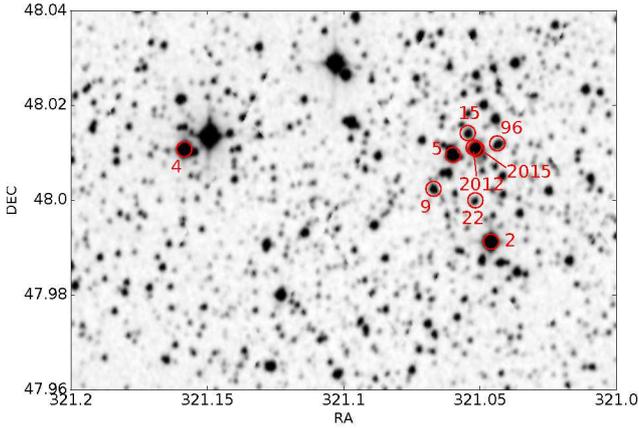}}
 \caption{Finding chart of the stars with available spectra. Number shows the ID according to WEBDA. See the correspondence with our ID in Table \ref{tab:specFP}.}\label{fig:FC2}
\end{figure}

All  the  observations  have  been  reduced,  following  standard  procedures  for  the  bias  subtraction,  flatfielding  and extraction,   using   the
\textsc{Starlink} software   packages \textsc{ccdpack} \citep{Draper00} and \textsc{figaro}  \citep{Shortridge97} and analysed using \textsc{figaro} and \textsc{dipso} \citep{Howarth97}.
 Sky was always subtracted by fitting a low-degree polynomial to points in two regions on each side of the spectra, as implemented by the \textsc{figaro}
routine \textsc{polysky}. 
See some of the obtained spectra in Fig.\,\ref{fig:spec}  and their finding chart in Fig.\,\ref{fig:FC2}.


\section{Results}\label{sec:Res}

\subsection{Photometric diagrams}\label{sec:photdiag}

\begin{figure}\centering
 \resizebox{\hsize}{!}{\includegraphics{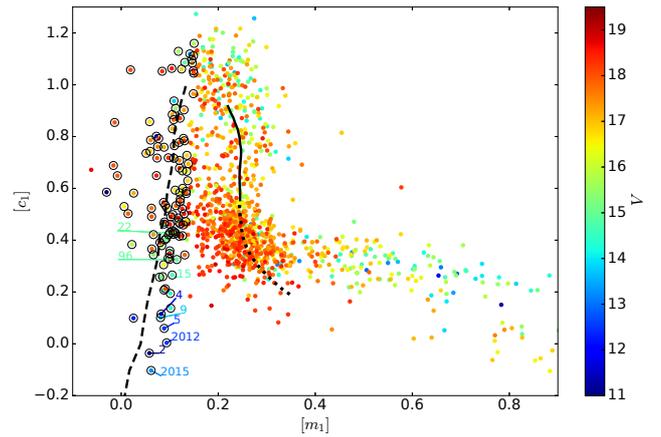}}
 \caption{ $[c_1]$--$[m_1]$ plot for all the available stars. Colour indicates the $V$ magnitude. The black lines show the standard relation from \citet{1987PASP...99.1184P}: 
dashed line for B stars, solid line for A stars, dotted line for F stars. Black circles indicate the likely B-stars (not necessarily members, see Sect.\,\ref{sec:membership}).
}\label{fig:c1m1}
\end{figure}

\begin{figure}\centering
 \resizebox{\hsize}{!}{\includegraphics{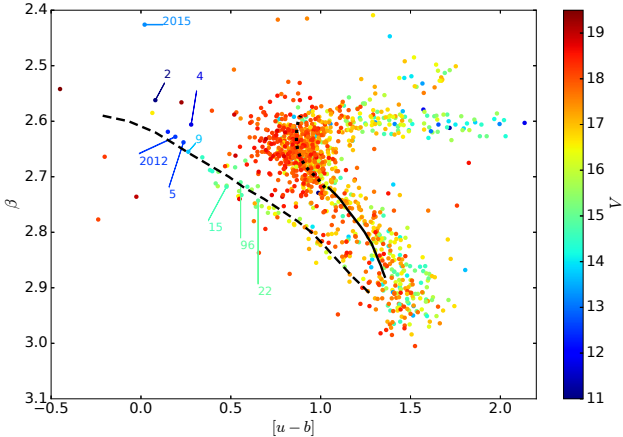}}
 \caption{ $\beta$--$[u-b]$ for all the available stars.  Colour indicates the $V$ magnitude. The black lines show the standard relation from \citet{1987PASP...99.1184P}: 
dashed line for B stars, solid line for A stars, dotted line for F stars. Stars with spectra are indicated following their ID$_{{\unit {WD}}}$.}\label{fig:Hbub}
\end{figure}

We compute the extinction-free indexes $[m_1], [c_1], [u-b]$, and $\beta$ using the standard values (i.e. assuming standard extinction laws where $E(m_1)=-0.33(b-y)$ and $E(c_1)=0.19E(b-y)$), that is:
\begin{eqnarray}\nonumber
 [m_1]&=&m_1+0.33\cdot(b-y),\\\nonumber
[c_1]&=&c_1-0.19\cdot(b-y), \\\nonumber
[u-b]&=&[c_1]+2\cdot[m_1],\\
\beta&=&H\beta_n-H\beta_w,
\end{eqnarray}
and compare them with the expected  sequences from \citet{1987PASP...99.1184P}. 
The $[c_1]$--$[m_1]$ plot (see Fig.\,\ref{fig:c1m1}) allows the identification of B-type stars, that lay on the left side of the plot, while the early-A stars are at the top, 
and later stars on the right part. 
Faint stars with larger photometric errors can blur the separation between early- and late-type stars, making difficult their classification. 
The same sequences are visible in the $\beta$--$[u-b]$ plot (see Fig.\,\ref{fig:Hbub}), with the early stars on the left part.
All the main sequence stars should lay around the \citet{1987PASP...99.1184P} sequences (except for emission line stars), independently of the distance and extinction.
Thus these plots provide a first estimation of the spectral type. 
However, these plots do not allow to 
distinguish between members and non-member stars.

\subsection{Spectral classification}\label{sec:spectra}

Spectral classification in the MK system is based on comparison with standard stars. 
The standards used are those listed in \cite{2004AN....325..380N}. 
The brightest star is LS\,III~$+47$\,36 (ID$_{{\unit {WD}}}$ =$\sharp$2, and $\sharp$786 in our catalogue) which demonstrates spectral variability. 
The spectrum shown in Fig.\,\ref{fig:spec}, corresponding to July 2002, shows double lines, indicative of two components (SB2). 
The brightest one is a (luminous) giant, with a spectral type not far from B2. The second component is almost as bright, given the line ratios. 
The likely detection of He\,\textsc{\lowercase{ii}}~4686\AA\ indicates that it is at least as early as B0.5, and so a B0.5\,V classification is quite possible. 
We note that  \citet{1965ApJS...12..215H}  classified the integrated spectrum as B5\,III, thus implying a much lower luminosity. 
LS\,III~$+47$\,37 is a visual pair (ID$_{{\unit {WD}}}$ =$\sharp$2012+$\sharp$2015, 
and $\sharp$779+$\sharp$782), resolved in our photometric catalogue, but not in the spectrum. Star ID$_{{\unit {WD}}}$=$\sharp$2012 is the brighter component. It is a Be star, with spectral type close to  B0.5\,Ve. 
Although we could detect the two stars, there could be some contamination in their photometric magnitudes. LS\,III~$+47$\,38 (ID$_{{\unit {WD}}}$=$\sharp$5, and $\sharp$771) is a B1\,IV-V star 
\citep[B2V according to][]{1965ApJS...12..215H}. LS\,III~$+47$\,39 (ID$_{{\unit {WD}}}$=$\sharp$4, and $\sharp$696) is clearly a giant, and we classify it as B1.5III, although \citet{1965ApJS...12..215H} 
classified it as a B0.5\,V.
Star ID$_{{\unit {WD}}}$=$\sharp$9 ($\sharp$763) is a B2\,V star. Stars ID$_{{\unit {WD}}}$=$\sharp$15, $\sharp$22 and $\sharp$96 ($\sharp$776, $\sharp$780, $\sharp$791 in our catalogue) are early- or mid-B stars. 
Their spectra are too noisy to give an accurate spectral type, but they all look close to B3\,V, and certainly not later than B5\,V. 
These spectral types are summarized in Table\,\ref{tab:specFP}.

\subsection{Membership and distance determination}\label{sec:membership}
\begin{figure}\centering
 \resizebox{\hsize}{!}{\includegraphics{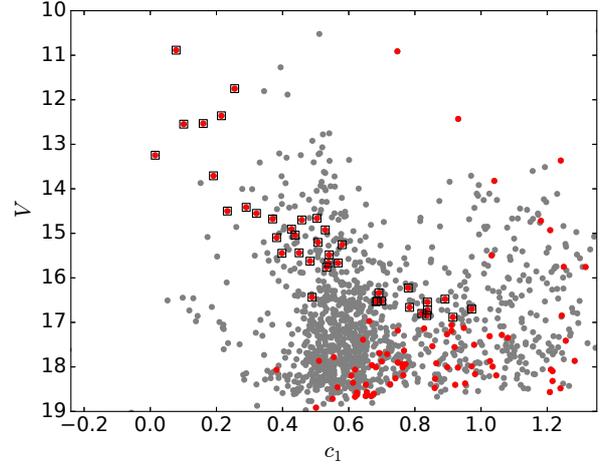}}
 \caption{ $V-c_1$ plot for the observed stars. Grey dots show all the stars. In red, those stars selected %in Fig.\,\ref{fig:c1m1B} 
as B type stars (even if they are not members, see Fig\,\ref{fig:c1m1}). Black squares show the B stars selected as likely members and used for absorption calculation (see Table \ref{tab:Bmemb})}\label{fig:Vc1B}
\end{figure}

\begin{figure}\centering
 \resizebox{\hsize}{!}{\includegraphics{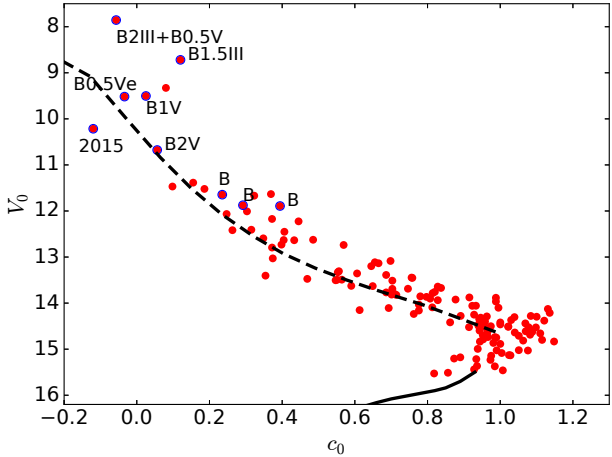}}
 \caption{ $V_0-c_0$ plot for all the stars that are likely members. The black lines show the standard relation from \citet{1987PASP...99.1184P}: dashed line for B stars, and solid line for A stars,
 with $V_0-M_V$=13.2\,mag. We indicate the spectral type for those stars with spectra available.}\label{fig:V0c0mod2}
\end{figure}

It is not a straightforward task to identify  the members of the cluster among all the stars observed in the same field of view. Several methods, including evaluation of kinematics
have been used for other authors to establish cluster membership. But this cluster is too far away to obtain good proper motions from the ground. 
Only \textit{Gaia}  will be able to give good proper motions for some of these stars in future data releases. 
In our case, we only have photometry, so we work with the colour-magnitude diagrams presented in Sect.\,\ref{sec:photdiag}.
As a first step we identified the B-type  stars present in the $[c_1]-[m_1]$ diagram (see Fig.\,\ref{fig:c1m1}). 
Among them, we selected as likely members, those located in the same sequence within the $V-c_1$ plot (i.e. see black squares in Fig.\,\ref{fig:Vc1B}). 
The B-type stars far from this sequence will have a complete different distance and absorption, so won't be located in the cluster.
For this first step, we reject any star from which we may have doubts that it is either a B star o a cluster member: 40 stars remain.
From the individual absorption computed for these 
stars using the \citet{2014A&A...568A.119M} method, we obtained their mean absorption  
(with a 3$\sigma$ clipping rejection):  $<A_V>$=$3.0\pm0.6$mag. The individual $A_V$ values for these stars can be found in Table \ref{tab:Bmemb}. 
We use this value to create the intrinsic $V_0-c_0$ diagram (Fig.\,\ref{fig:V0c0mod2}),  whose values are also in Table \ref{tab:Bmemb}. Then, we can fit the sequence and develop a more accurate cluster member selection (for stars beyond the B-type already selected). 
A star will be considered as likely member if it lies on the sequence in the $V_0-c_0$ plot, and at the same time, the spectral type matches in the $[c_1]-[m_1]$ and $\beta-[u-b]$ plots.
The 139 stars selected as likely members can be found in Fig.\,\ref{fig:V0c0mod2} and indicated in the Table \ref{tab:TablePhot}. The distance modulus fitted is $|V_0-M_V|$=13.2$\pm$0.2 (the error indicates the uncertainty in positioning 
the theoretical ZAMS and its identification as a lower envelope). This leads to a distance to the cluster of 4.4$\pm0.4$\,kpc.
Our observations reach deep enough to detect  early-A type stars in the cluster. That is very useful since the fit of the turning point in the $V_0-c_0$ diagram (Fig.\,\ref{fig:V0c0mod2})  allows us to obtain a reliable distance.

\subsection{Physical parameters for individual stars}\label{sec:FP}

We use the method described in \citet{2014A&A...568A.119M} to compute individual physical parameters for hot stars with $T_{\unit{eff}}>7000\,\unit{K}$. 
This method is based on atmospheric models and evolutionary 
tracks and it provides individual physical parameters such as distances, absorptions, and effective temperatures ($T_{\unit{eff}}$). 
This method is good for large samples of stars, and to treat them statistically  \citep[as has been shown in][]{2015A&A...577A.142M}. However, one should be careful when dealing with individual stars, as those 
values may have  large individual errors.
In Table \ref{tab:specFP} we can see, as an example, some physical parameters obtained for the stars with spectra available discussed in previous section. 
 $T_{\unit{eff}}$, $\log g$, distance, $M_V$, $A_V$ are available for the 515 stars with $T_{\unit{eff}}>7000\,\unit{K}$ (not necessarily members) and can be found in Table \ref{tab:TablePF}. 
 The errors presented for those physical parameters have been computed through Monte Carlo simulations, assuming random Gaussian errors for the input photometry.
As we discussed in Sect.\,\ref{sec:Obs}, the photometric errors can be slightly underestimated when all the observations have been taken consecutively. In this case, the errors in physical parameters 
might be underestimated too.

The distance and absorption computed for the cluster are more accurate when we use the colour-magnitude diagram with all the 
stars. Nevertheless, we can also compute the mean values for the 139 cluster members with physical parameters available.
We computed the median with a 3$\sigma$ clipping, obtaining  4.0$\pm$0.8\,kpc, which is smaller but compatible with the value obtained from the fitting.
The mean absorption for all the member stars is $<$$A_V$$>$=3.1$\pm$0.6\,mag (to be compared with the $<$$A_V$$>$=3.0$\pm$0.6\,mag when we only used the B-type stars), 
and their mean distance modulus: $<V_0-M_V>$=13.0$\pm$0.6\,mag (where the errors are computed as the standard deviation). 
The computed mean absorption has a large dispersion, since some 
of the stars have a computed high individual absorption. However, some of the stars with larger absorption are also faint, and so have larger errors. 
Actually, all the stars selected as members, and with $A_V>$3.8 are all fainter than $V$=16.5. 

\begin{table*}
 \caption{List of stars with spectra available. ID$_{{\unit {WD}}}$ is the identifier according to WEBDA. ID is the internal identifier from our catalog.
SP is the spectral type obtained from spectra. Str\"omgren photometry and computed physical parameters are shown.
 The effective temperature, gravity, and visual absorption were 
computed according to \citet{2014A&A...568A.119M}. }\label{tab:specFP}
\centering\begin{tabular}{lcc|ccccccccccc}
Name&ID$_{{\unit {WD}}}$  &ID	&SP&RA&DEC&$V$&$(b-y)$&$m_1$&$c_1$&$\beta$	 &$T_{\unit{eff}}$(K) & $\log g$    &$A_V$(mag) \\ \hline
LS\,III~$+47$\,39&4    &696	&B1.5III  &21 24 38.0& +48 00 39	&11.75  	 &0.74  &0.26 & 0.58  &  2.61  &21030  &3.57       &3.55 \\
                 &9    &763	&B2V      &21 24 16.1& +48 00 09	&13.71  	 &0.47  &0.19 & 0.40  &  2.66  &21830  &4.48        &2.46 \\
LS\,III~$+47$\,38&5    &771	&B1V      &21 24 14.4& +48 00 35	&12.53  	 &0.53  &0.16 & 0.44  &  2.64  &22910  &4.79        &2.71 \\
                 &15   &776	&mid B  &21 24 13.1& +48 00 51	&14.68  	 &0.55  &0.37 & 0.47  &  2.72  &17110  &4.71        &2.68 \\
LS\,III~$+47$\,37a&2012 &779	&B0.5Ve   &21 24 12.5& +48 00 40	&12.55  	 &0.51  &0.10 & 0.43  &  2.63  &25060  &4.46        &2.65 \\
                 &22   &780	&mid B        &21 24 12.4& +48 00 00	&14.92  	 &0.54  &0.53 & 0.47  &  2.74  &14670  &4.46        &2.57 \\
LS\,III~$+47$\,37b&2015 &782	&?	  &21 24 12.3& +48 00 39	&13.24  	 &0.62  &0.02 & 0.48  &  2.43  &25990  &3.00     &3.04 \\
LS\,III~$+47$\,36&2    &786	&\tiny{B2III+B0.5V}   &21 24 11.0& +47 59 29	&10.89   &0.60  &0.08 & 0.46  &  2.56  &25910  &4.43      &3.05 \\
                 &96   &791	&mid B  &21 24 10.5& +48 00 43	&14.91  	 &0.54  &0.43 & 0.47  &  2.71  &16270  &4.41      &2.60   \\
\end{tabular}
\end{table*}

\subsection{Additional data}\label{sec:other}
\begin{figure}\centering
 \resizebox{\hsize}{!}{\includegraphics{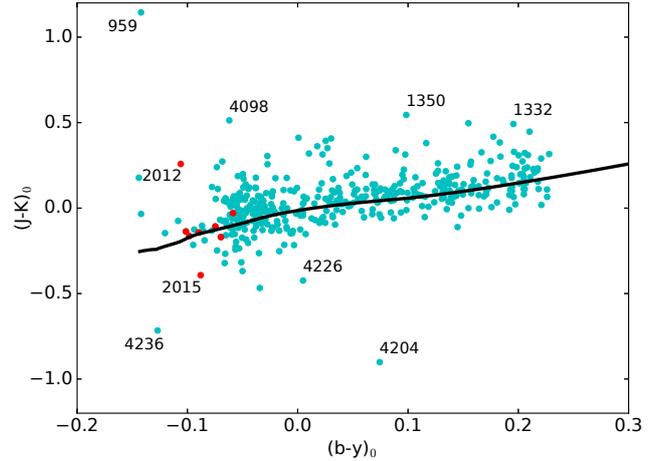}}
 \caption{ $(J-K)_0$ vs. $(b-y)_0$. The black line shows the expected main sequence obtained by combining grids of the ATLAS9 model atmosphere from \citet{2004astro.ph..5087C,2006A&A...454..333C} and \citet{1998A&A...333..231B}. 
 In red, stars with available spectra. Stars with the ID labeled are discussed in the text.}\label{fig:2mass}
\end{figure}

We provide Str\"{o}mgren magnitudes and photometric indexes for up to 1233 stars in the field. In addition, 1186 of these stars have infrared photometry available from the 2MASS  catalog \citep{2006AJ....131.1163S}, 
although only 1034 of those have the full set of $JHK$ magnitudes (rejecting those that are upper limits).
We compare here these two sets of photometry in order to check consistency and look for outliers. 
In Set.\,\ref{sec:FP} we obtained individual absorptions in the visual band ($A_V$) from the Str\"omgren photometry for stars with $T_{\unit{eff}}>$7000\,K. 
In this section,  we transform those values to the infrared bands following the expressions by 
\citet{2004AJ....128.2144M} that relate absorption at different wavelengths. 
Using the absorptions in the different bands, we obtained the intrinsic colours $(b-y)_0$ and $(J-K)_0$.
Only 423 stars fulfill the requirement of having both the $JHK$ magnitudes and the individual physical parameters (i.e. $T_{\unit{eff}}>$7000\,K) needed to develop this comparison.
Errors in 2MASS for these stars have mean values of $<\epsilon_J>=0.05$\,mag, $<\epsilon_H>=0.05$\,mag, and $<\epsilon_K>=0.08)$ and can reach values of 0.1\,mag in $\epsilon_J$,  0.2 in $\epsilon_H$ and 0.2 in $\epsilon_K$.
Figure\,\ref{fig:2mass} shows the intrinsic $(J-K)_0$ vs. $(b-y)_0$  colours for these 423 stars. 
This has also been compared with the theoretical main-sequence colours --\citet{1998A&A...333..231B} for $JHK$, and \citet{2004astro.ph..5087C,2006A&A...454..333C} for $uvby$. 
As expected, most of the stars follow this sequence. Up to 85\% of the stars are closer than 0.1\,mag to the sequence. The stars that are far from the sequence are usually non-main sequence stars, 
emission line stars, or stars with wrong computed absorption (due to e.g. large photometric errors). In particular we note star ID$_{{\unit {WD}}}$=$\sharp$2012, 
that is known to be emission line, and star ID$_{{\unit {WD}}}$=$\sharp$2015 shown to have a strange behaviour in 
colour-colour plots due to contamination. Star ID=959 has $V$=17.16, with very strong emission ($r-H\alpha$=1.58, $\beta$=1.58) and its photometry indicates that is a very hot and distant star.
On the other hand, stars ID=1332, 1350, 4098, 4204, 4226, and 4236 are all fainter than $V$=18\,mag, and their location in the diagram is due to large photometric errors either in Str\"{o}mgren and/or 2MASS.

\begin{figure}\centering
 \resizebox{\hsize}{!}{\includegraphics{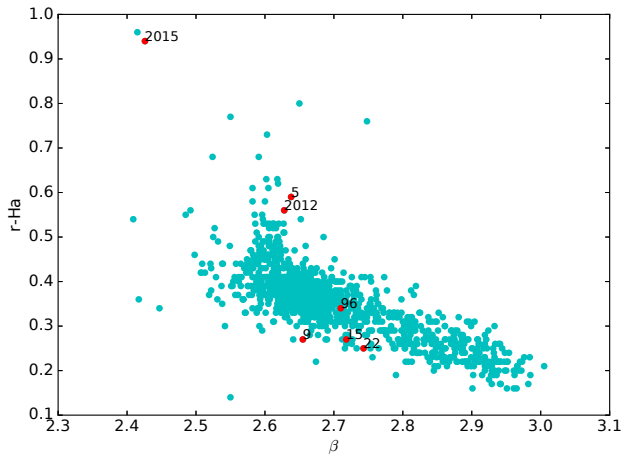}}
 \caption{Combination of IPHAS and Str\"omgren data: $(r-H\alpha)$ vs. $\beta$. Stars with available spectra are indicated in red and labeled according to WEBDA ID.}\label{fig:iphas}
\end{figure}

Emission line stars can be detected using IPHAS data \citep{2014MNRAS.444.3230B}.
Almost all the stars in the field (up to 1220) have data in the three bands $i, r$, and $H\alpha$.
Figure\,\ref{fig:iphas} shows some stars above the expected sequence, such as ID$_{{\unit {WD}}}$=$\sharp$5, and $\sharp$2012. 
Star ID$_{{\unit {WD}}}$=$\sharp$2012 is already a known emission line star.
The star ID$_{{\unit {WD}}}$=$\sharp$5 is also above the expected sequence, although its spectrum does not show emission lines. 
Actually this star has a flag in the IPHAS catalog that indicates saturation (it is $V$=12.5), which gives it non-reliable IPHAS photometry.
IDs from both IPHAS and 2MASS resulting from the crossmatch with our catalog are available in Table \ref{tab:TablePhot}.


\subsection{Cluster geometry}\label{sec:density}

\begin{figure}\centering
\resizebox{\hsize}{!}{\includegraphics{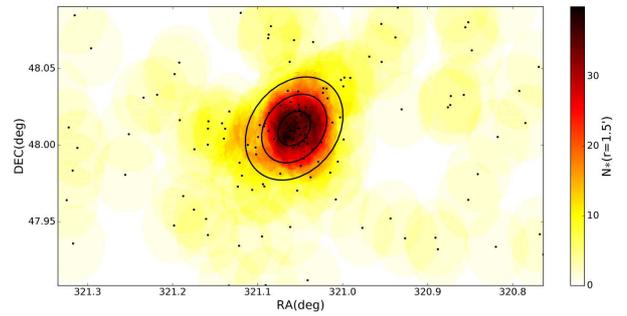}}
 \caption{Sky distribution of the 139 stars likely members (black dots) and density computed as the number of stars closer than 1$\farcm$5 to each point. 
The black solid lines show the fitted 2D gaussian at 1, 2 and 3$\sigma$.}\label{fig:3Ddens}
\end{figure}

Figure \ref{fig:3Ddens} shows the sky density distribution of the 139 likely member stars in the observed region around the cluster. The density at each direction of the sky
has been computed as the number of stars located closer than 1$\farcm$5.
We clearly detect the overdensity at the central part of the cluster. 
We use these data to fit a 2-D gaussian to the sky density of the members, such as:
\begin{equation}
 n=n_0+n_{\unit{max}} \cdot \exp ^{- (A(\alpha-\alpha_0)^2 + 2B(\alpha-\alpha_0)(\delta-\delta_0) + C(\delta-\delta_0)^2))}
\end{equation}
with
\begin{eqnarray}
    A &=& \frac{\cos^2\theta}{2\sigma_{\alpha}^2} + \frac{\sin^2\theta}{2\sigma_{\delta}^2}\\\nonumber
    B &=& \frac{-\sin(2\theta)}{4\sigma_{\alpha}^2} + \frac{\sin(2\theta)}{4\sigma_{\delta}^2}\\\nonumber
    C &=& \frac{\sin^2\theta}{2\sigma_{\alpha}^2} + \frac{\cos^2\theta}{2\sigma_{\delta}^2}
\end{eqnarray}
where the seven parameters, obtained through a least squares fit to the sky star density, are:
1)the offset due to the zero point $n_0$=963$\pm$4 $\star/\circ^2$, 
2)the amplitude of the gaussian, i.e. the peak density $n_{\unit{max}}$=18020$\pm$30 $\star/\circ^2$, 
3)the right ascension of the centre of the cluster $\alpha_0$=$321.05703\pm0.00007^{\circ}$,
4)the declination of the centre of the cluster $\delta_0$=$48.01089\pm0.00004^{\circ}$,
5)the sigma in the right ascension direction $\sigma_{\alpha}$=$0.03890\pm0.00007^{\circ}$, i.e. 2$\farcm$3, % $\arcmin$,
6)the sigma, in the declination direction $\sigma_{\delta}$=$0.02141\pm0.00004^{\circ}$, i.e. 1$\farcm$3, and
7)the position angle of the ellipse $\theta$=$11.90\pm0.02^{\circ}$ with respect to the vertical.
The centre of the cluster in galactic coordinates is $(\ell,b)$=$(91.19817^{\circ},-1.67647^{\circ})$. That is 34$\arcsec$ from the value given by \cite{2013A&A...558A..53K}.
The results provide an elliptic profile with an eccentricity:
\begin{equation}
\epsilon=\sqrt {1-\frac{(\sigma_{\alpha} \cos\delta_0)^2}{\sigma_{\delta}^2}}=0.57. 
\end{equation}

Considering that almost all the stars of the cluster should be contained within 3$\sigma$, we can establish the radial size of the cluster like: $r_C=3\sqrt{(\sigma_{\alpha} \cos\delta_0)^2+\sigma_{\delta}^2}=6\farcm$1.
That would translate into a physical radius of 7.8\,pc (since the cluster is at 4.4\,kpc).

\subsection{Age and mass determinations}\label{sec:Sectage}
\begin{figure}\centering
 \resizebox{\hsize}{!}{\includegraphics{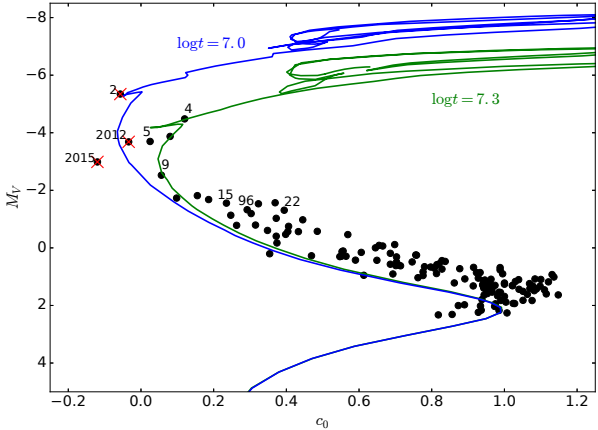}}
 \caption{Colour magnitude diagram   $M_V-c_0$ for the stars selected as a possible members. Two isochrones from \citet{2008A&A...482..883M} are overplotted with 
$\log_{10} (t(\unit{yr}))$=7.0 (blue) and 7.3 (green). Three stars with unreliable physical parameters are indicated with a red cross (see text for discussion).}\label{fig:age}
\end{figure}

The cluster studied here is  very young. Since only two giants (ID$_{{\unit {WD}}}$=$\sharp$2, and $\sharp$4) were found in this cluster (and $\sharp$2 cannot be used because it has contamination from the binary), 
the isochrone fitting is quite difficult. 
In Fig.\,\ref{fig:age} we see the $M_V-c_0$ diagram, 
with some isochrones overplotted. We see that most of the stars are well located on the main sequence. 
We should remove from the analysis the stars ID$_{{\unit {WD}}}$=$\sharp$2015 and $\sharp$2012 since we checked they have contamination from the other, plus $\sharp$2012 is an emission line star. 
The giant star ID$_{{\unit {WD}}}$=$\sharp$2 has been proved to be a binary (see Sect.\,\ref{sec:spectra}, 
so its photometric results are not reliable either.
Once these are removed, the remaining stars 
fit very well with the 
isochrone of $\log_{10} (t(\unit{yr}))$=$7.3$ (20\,Myr). Specially the star $\sharp$4, which is a giant B1.5III. 
However, this result is based in only few stars, that are also not too evolved in the giant branch. 
In addition, the presence of two B0.5V stars, suggests that the age cannot be much older than 10-12Myr.
That leads to the conclusion that the age is very difficult to establish from the photometric diagrams, but in any case we can set an upper limit of $\log_{10} (t(\unit{yr}))$=$7.3$.

\begin{figure}\centering
 \resizebox{\hsize}{!}{\includegraphics{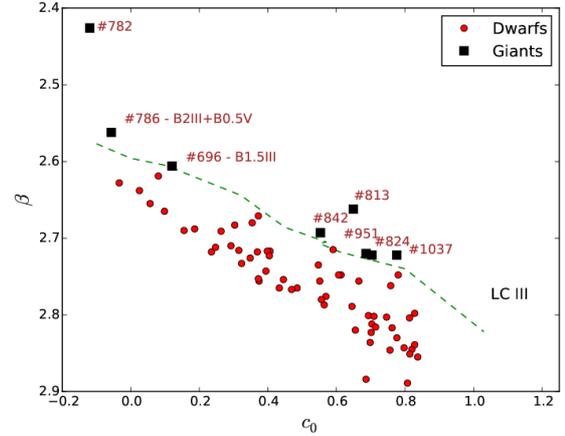}}
 \caption{$\beta$ vs. $c_0$ distribution for the B-stars selected as members, and used for the mass derivation. The green line indicates the luminosity class III sequence according to \citep{1978AJ.....83...48C}.
 Red dots indicate the stars classified as dwarfs while black  squares show those classified as giants (with our catalog ID indicated). }\label{fig:mass}
\end{figure}

In order to estimate the cluster mass we used  the multiple-part power law Initial Mass Function (IMF) defined by
\citep{2001MNRAS.322..231K}. We set the free parameter of the IMF by
counting the stars within a certain mass range. Because of the completeness of
our photometry (at least down to A1\,V) we choose the entire B-range (i.e.
B0\,V\,--B9\,V) with masses comprised between 19.95 and 2.57\,$\unit{M}_{\sun}$
according to the calibration of \citep{1992msp..book.....S}.
We rejected from this sample the giant stars, that are separated from the dwarfs using the different luminosity class sequences provided by \citep{1978AJ.....83...48C}.
As seen in Fig.\,\ref{fig:mass}, the two known giant stars with spectroscopy (ID$_{{\unit {WD}}}$=$\sharp$4 and ID$_{{\unit {WD}}}$=$\sharp$2) are well separated from the dwarfs. The only star far from the sequence is
the already discussed binary ID$_{{\unit {WD}}}$=$\sharp$2015+$\sharp$2012. The other stars located above the LCIII sequence (ID= 813, 824, 842, 951, 1037) are probably Be stars 
(all of them have $r$-H$\alpha>$0.33, in the upper range for B-type stars) or background stars wrongly assigned as members. In any case, they are few stars and do not affect the computed mass of the cluster.
In the total area observed, we obtained 60\,$\pm$\,8 (poissonian uncertainty) B-dwarf stars.
Then,
by integrating the IMF, we derived a present-day cluster mass of 1260\,$\pm$\,160\,$\unit{M}_{\sun}$, which would correspond to an initial one of the cluster of around 1600\,$\unit{M}_{\sun}$.

\section{The connection with the Perseus spiral arm}\label{sec:disc}
The location of this cluster makes us consider its relation with the Perseus spiral arm, within the framework of the spiral arm structure of the Milky Way.
The understanding of the nature, origin and mechanisms that drive the spiral arms in our Galaxy is still a fundamental problem in astrophysics. 
Even the properties and the extent of the Local Arm are still under debate \citep{2016arXiv161000242X}. 
It is well assumed that the spiral 
arms are an important mechanism that affects the dynamics and evolution of the Milky Way. However, there are still lots of open questions. 
Several physical processes have been suggested for the  origin and 
supporting mechanism of this structure. The most important are: 
a) The spiral structure is tidally induced, either through external \citep[with a companion galaxy,][]{2008ASPC..396..321D}
 or internal \citep[driven by a bar]{2010ApJ...715L..56S,2009MNRAS.394...67A} interactions,
b) The spiral structure is the result of quasi-stable global modes in the disc, as predicted by the density wave theory \citep{1964ApJ...140..646L},
c) The spiral structure is produced by local, self-gravitational instabilities \citep{1977ARA&A..15..437T}, and
d) The spirals are manifold-driven \citep{2007A&A...472...63R,2012MNRAS.426L..46A}.
These theories are not mutually exclusive, and the spiral structure of different galaxies might be dominated by different mechanisms. In any case, they are tightly related with star formation, since each of
 the mechanisms predicts different temporal patterns of star formation. That leads to different age trends along or across the arms, that will be reflected in the distribution in ages of the tracers.
The analysis of this distribution can thus unveil the underlying cause of the spiral structure.
Indeed, \citet{2014ApJS..215....1V} suggests that different observables may locate the spiral arms at slightly different locations. 
This author found that the gas and dust concentrate at the inner part of the arms, with the density peak due to stars at slightly further positions.
This distribution matches with the results discussed by 
\citet{2015A&A...577A.142M}, that found a dust layer at the inner part of the stellar density wave of the Perseus arm.
Another work supporting density-wave theory, in this case for external galaxies, is \citet{2016ApJ...827L...2P}
that used observations at different wavelengths to state that pitch angle is varying for different galactic components, such gas, star forming regions and older stars.

\begin{figure}\centering
 \resizebox{\hsize}{!}{\includegraphics{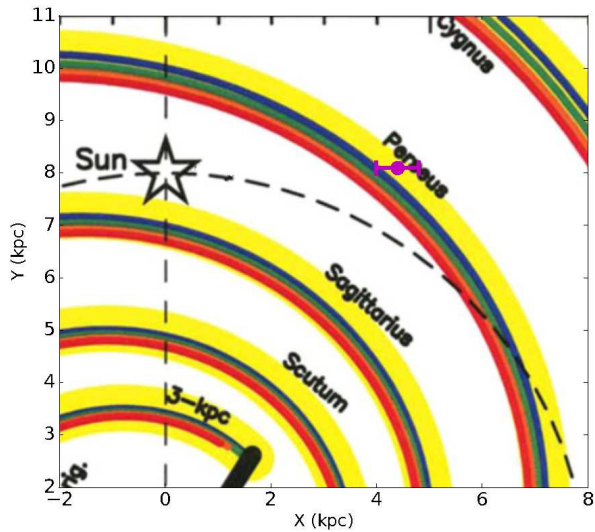}}
 \caption{The background figure is the structure of the spiral arms suggested by \citet{2014ApJS..215....1V}, where each colour indicates a different tracer (see his Fig.\,1).  Stars
are plotted in yellow, 12CO (and H2) molecules are in blue, hot dust in red, thermal and relativistic electrons and HI atoms in green, and cold dust with FIR cooling lines
in orange.
We overplotted the expected location of NGC 7067 in magenta, with the corresponding error bars.}\label{fig:ambVallee}
\end{figure}

The obtained distance to NGC 7067 and its direction, allow us to place the cluster within the Perseus spiral arm 
\citep[see Fig.\,\ref{fig:ambVallee} where we overplotted the location of the cluster in the pattern of the spiral arms suggested by][]{2014ApJS..215....1V}.
According to this image, the cluster would be slightly shifted
to external positions of the arm, where stars would be located, and slightly further away than the gas and dust, that peak at the internal part of the arm.
We must mention that this pattern is based on tangencies to the inner arms of the Galaxy, and the 
extrapolation to the external arms, together with the uncertainty of the obtained photometric distance to the cluster,
 may not be sufficient to establish the exact relative location between the arm and the cluster.  
However, this result could be understood as a movement of the cluster from the internal part of the arm, where star formation would occur, to external parts.
\citet{1969ApJ...158..123R} analysed the star formation in the
 shocks produced by the spiral arms, in the framework of density wave theory, and located the region of newly born stars (up to 30\,Myr, $\log_{10} (t(\unit{yr}))$=7.5) very close to the potential minimum. 
Other spiral arm theories predict different distributions. 
In this context, \citet{2010MNRAS.409..396D} developed some simulations of the age distribution of clusters born in the spiral arms, obtaining 
different age dispersions depending on the mechanism that is supporting the spiral arm. 
A single cluster with the given accuracy in distance and age is not enough to distinguish among these theories, 
but the current result opens new options for the analysis of the distribution of the spiral arms of the Galaxy.

The Perseus arm in the second quadrant is optimal to study the distribution of open cluster ages in a galactic region wide enough (90$\degr<l<140\degr$) to cover a substantial part of a spiral arm.
And it is in the outer part of the disc where cluster populations are easier 
to detect and lower interstellar absorption allows us to reach further distances. 
It is a well defined structure, rich in open clusters, with more than 50 detected so far. 
 The current data available in the literature 
 come from very  inhomogeneous datasets, and cannot lead to any firm conclusion. 
Also the different isochrones used to derive ages for each of the references, may lead to different results and inaccuracies that should be avoided.
In this direction, much more reliable and homogeneous data and methods to derive physical parameters are needed to reach conclusive results.
Str\"omgren $uvby\beta$ photometry is optimal to undertake this study since it is the most appropriate photometric system to study young stars and obtaining accurate
estimates of individual distances and ages. These photometric estimates,  combined with spectra for some of the stars in the cluster should provide much more reliable results.
Absolute distances from the \textit{Gaia} mission \citep{2016arXiv160904153G}, once combined with the photometric relative ages, will ensure accuracy enough 
to trace any possible trend in ages in the spiral arms.

\section{Conclusions}\label{sec:conclusions}

The analysis of the Str\"omgren data, together with the nine available spectra, allowed us to compute the physical parameters for the cluster NGC 7067.
 139 likely member stars where selected from the photometric diagrams. From a main sequence fitting to the cluster data we obtained a distance of 4.4$\pm0.4$\,kpc, 
 while using the mean of individual star distances we obtained 4.0$\pm$0.8\,kpc.
Previous references gave distances in the range [2.3-4.5]\,kpc. For example,
\citet[][]{1961LowOB...5..133J}, obtained 2900\,pc using 29 stars
\citet[][]{1963ZA.....57..117B}, found 3950\,pc using the same sample
\citet[][]{1965MmSAI..36..283B}, computed 4500\,pc, adding 60 stars more to the sample, and 
\citet[][]{1973A&AS....9..261H}, derived two values, 3963 and 4406\,pc, the first one using 111 stars and the second one using the same previous subset of 29 stars.
All these studies, used few stars and only broad band $UBV$ photometry.
\citet{2004MNRAS.349.1481Y} were the first using CCD photometry in this cluster increasing the sample up to $\sim$2000 stars in the field, and using also infrared filters. 
They gave an age of $\log_{10} (t(\unit{yr}))$=8, that is not compatible with the presence of early-B stars in the cluster (which is now reliable thanks to the available spectra). 
That may be partially explained by the fact that they provided a 
closer photometric distance of 3600\,pc. The brighter stars in the clusters ($V\sim 12.5$) would have an absolute magnitude of $M_V$=-2.6 using their cluster parameters, 
while they would have $M_V$=-3.7 using the distance and absorptions obtained in our study. 
So,  even with the cluster parameters that they derive,  the brightest stars would have magnitudes typical of early-type stars and thus incompatible with the age that they find. 
% So it would still be early type stars not compatible with such an old age.
The fitted upper limit on the age of $\log_{10} (t(\unit{yr}))\lesssim7.3$ ($\sim$20Myr) is in agreement with the value of $\log_{10} (t(\unit{yr}))$=7.1 published by \citet{1968ArA.....5....1L}.
The derived present mass of 1260\,$\pm$\,160\,$\unit{M}_{\sun}$ is clearly larger than the 550\,$\unit{M}_{\sun}$ obtained by \citet{1983A&A...121..237B}. 
The main differences between the two estimations are different IMF used (\citet{2001MNRAS.322..231K} vs. \citet{1955ApJ...121..161S}), and that while we integrated the mass directly 
from our Str\"omgren photometry, they transformed from a relative mass \citep[normalized to Trumpler 1;][]{1978stfo.book.....R} using only three clusters to compute the zero points.

We also provided a new value for the centre of the cluster in J2000 of  $(\alpha,\delta)$=$(21\colon24\colon13.69,+48\colon00\colon39.2)$, or $(\ell,b)$=$(91.198167\degr,-1.676472\degr)$, %91\colon11\colon53.4,-01\colon40\colon35.3)$, 
with a size radius of $r_C$=$6\farcm1$. 
Individual physical parameters for hot stars have been also computed using the method from \citet{2014A&A...568A.119M} and their mean values of distance and absorptions are in agreement with the results of sequence fitting.
We complemented our photometry with data from 2MASS and IPHAS. The first catalog was used to compare and detect outliers while IPHAS catalog was used to detect emission line stars.

Finally, we discussed the relation of the cluster with the Perseus spiral arm. Its age and location suggest that it was formed in the spiral arm,
making it a perfect candidate for the analysis of the spiral arm theory in the Perseus spiral arm.

\section*{Acknowledgements}
Based on observations made with the Isaac Newton Telescope operated on the island of La Palma by the Isaac Newton Group in the Spanish Observatorio del Roque de los Muchachos of the 
Instituto de Astrof\'isica de Canarias, and with the 1.93-m telescope at Observatoire de Haute-Provence (CNRS), France.
MM, IN, AM, and JA, work is partially supported  by the Spanish Government Ministerio de Econom\'{\i}a y Competitividad under grant  AYA2015-68012-C2-2-P (MINECO/FEDER).
AM acknowledges support from the Generalitat Valenciana through the grant BEST/2015/242 and the Ministerio de Educaci\'{o}n Cultura y Deporte through the grant PRX 15/00030. 
FV acknowledges support by the Spanish Ministry of Economy and Competitiveness (MINECO) through grant ESP2014-57495-C2-2-R.
LC, FF, CJ, MRG   are   supported   by the MINECO (Spanish Ministry of Economy) --FEDER through grant ESP2013-48318-C2-1-R and ESP2014-55996-C2-1-R 
and MDM-2014-0369 of ICCUB (Unidad de Excelencia 'Mar\'{\i}a de Maeztu'). MTC and EJA are supported by the project AYA2013-40611-P.
This publication makes use of data products from the Two Micron All Sky Survey, which is a joint project of the University of Massachusetts and the Infrared Processing
and Analysis Center/ California Institute of Technology, funded by the National Aeronautics and Space Administration and the National Science Foundation. 
This paper makes use of data obtained as part of the INT Photometric H$\alpha$ Survey of the Northern Galactic Plane (IPHAS, www.iphas.org) carried out at the Isaac Newton Telescope. 
All IPHAS data are processed by the Cambridge Astronomical Survey Unit, at the Institute of Astronomy in Cambridge. 
The band merged DR2 catalogue was assembled at the Centre for Astrophysics Research, University of Hertfordshire, supported by STFC grant ST/J001333/1. 
This publication makes use of data products from the Digitized Sky Surveys that were produced at the Space Telescope Science Institute under U.S. Government grant NAG W-2166. 
The images of these surveys are based on photographic data obtained using the Oschin Schmidt Telescope on Palomar Mountain and the UK Schmidt Telescope. 
The plates were processed into the present compressed digital form with the permission of these institutions.

%%%%%%%%%%%%%%%%%%%%%%%%%%%%%%%%%%%%%%%%%%%%%%%%%%



\bibliographystyle{aa}
\bibliography{NGC7067-rev}

%%%%%%%%%%%%%%%%%%%%%%%%%%%%%%%%%%%%%%%%%%%%%%%%%%

%%%%%%%%%%%%%%%%% APPENDICES %%%%%%%%%%%%%%%%%%%%%

\appendix

\section{Tables}\label{sec:AppTables}
All the available data for the individual stars are available online. Here we present the ten first rows for each of them. 
Table \ref{tab:TablePhot} provides the photometry for the 1233 stars, as well as membership determination. 
Table \ref{tab:Bmemb} shows the parameters for the B stars classified as likely members.
Table \ref{tab:TablePF} shows the computed physical parameters for 515 hot stars.
% \clearpage 
\onecolumn

\begin{landscape}
\begin{longtable}{cccccccccccccccc}
\caption{Coordinates and photometric parameters for the 1233 stars available. Last column is zero for non-members stars and 1 for likely members. 
Crossmatched IDs with IPHAS and 2MASS catalogues are also shown. Only ten rows are presented here. Full table available online.}\\\label{tab:TablePhot}
% \centering\begin{tabular}{cccccccccccccc}
ID&RA&DEC&V&$\varepsilon_V$ &$(b-y$)&$\varepsilon_{(b-y)}$&$m_1$& $\varepsilon_{m_1}$ &$c_1$& $\varepsilon_{c_1}$ &$\beta$&$\varepsilon_{\beta}$& member& ID IPHAS & ID 2MASS\\\hline
%   623   &21:25:16.1&+48:03:41&  15.91&  0.02 & 0.79& 0.01  &-0.01& 0.03 & 0.55& 0.01 & 2.67 & 0.01 &0\\
%   624   &21:25:15.7&+48:01:08&  14.52&  0.01 & 1.18& 0.02  & 0.13& 0.02 & 0.59& 0.01 & 2.62 & 0.01 &0\\
%   625   &21:25:15.4&+47:56:07&  13.51&  0.01 & 0.60& 0.01  & 0.02& 0.02 & 1.19& 0.01 & 2.93 & 0.01 &0\\
%   626   &21:25:15.0&+47:57:50&  16.28&  0.01 & 1.03& 0.01  &-0.12& 0.02 & 0.62& 0.02 & 2.66 & 0.01 &0\\
%   627   &21:25:12.1&+47:58:36&  15.69&  0.01 & 0.94& 0.01  &-0.09& 0.01 & 0.77& 0.01 & 2.69 & 0.01 &0\\
%   628   &21:25:12.0&+48:02:37&  14.67&  0.01 & 0.78& 0.01  &-0.06& 0.02 & 1.09& 0.01 & 2.77 & 0.01 &0\\
%   629   &21:25:11.5&+48:01:20&  14.58&  0.01 & 0.55& 0.01  & 0.01& 0.01 & 1.17& 0.01 & 2.96 & 0.01 &0\\
%   630   &21:25:10.7&+48:03:16&  14.43&  0.01 & 0.74& 0.02  & 0.06& 0.02 & 0.98& 0.01 & 2.84 & 0.01 &0\\
%   631   &21:25:10.6&+48:02:50&  15.17&  0.01 & 1.01& 0.01  &-0.09& 0.02 & 0.64& 0.01 & 2.66 & 0.01 &0\\
%   632   &21:25:10.3&+47:56:32&  15.35&  0.02 & 1.22& 0.01  & 0.07& 0.03 & 0.57& 0.01 & 2.62 & 0.01 &0\\
  
623   &21:25:16.1  &+48:03:41 &15.91  &0.01  &0.79  &0.01 &-0.01  &0.03  &0.55  &0.01  &2.67  &0.01 &0 &J212516.10+480341.5 &21251609+4803414\\
624   &21:25:15.7  &+48:01:08 &14.52  &0.01  &1.18  &0.02 & 0.13  &0.02  &0.59  &0.01  &2.62  &0.00 &0 &J212515.73+480107.3 &21251572+4801073\\
625   &21:25:15.4  &+47:56:07 &13.51  &0.01  &0.60  &0.01 & 0.02  &0.01  &1.19  &0.01  &2.93  &0.01 &0 &J212515.42+475606.9 &21251539+4756068\\
626   &21:25:15.0  &+47:57:50 &16.28  &0.01  &1.03  &0.01 &-0.12  &0.02  &0.62  &0.02  &2.66  &0.01 &0 &J212515.02+475749.5 &21251500+4757494\\
627   &21:25:12.1  &+47:58:36 &15.69  &0.01  &0.94  &0.01 &-0.09  &0.01  &0.77  &0.01  &2.69  &0.01 &0 &J212512.17+475835.8 &21251215+4758357\\
628   &21:25:12.0  &+48:02:37 &14.67  &0.01  &0.78  &0.01 &-0.06  &0.02  &1.09  &0.01  &2.77  &0.01 &0 &J212512.01+480236.6 &21251199+4802365\\
629   &21:25:11.5  &+48:01:20 &14.58  &0.01  &0.55  &0.01 & 0.01  &0.01  &1.17  &0.01  &2.96  &0.00 &0 &J212511.54+480119.5 &21251153+4801195\\
630   &21:25:10.7  &+48:03:16 &14.43  &0.01  &0.74  &0.01 & 0.06  &0.02  &0.98  &0.00  &2.84  &0.00 &0 &J212510.66+480316.1 &21251064+4803161\\
631   &21:25:10.6  &+48:02:50 &15.17  &0.01  &1.01  &0.01 &-0.09  &0.02  &0.64  &0.01  &2.66  &0.00 &0 &J212510.58+480249.6 &21251057+4802495\\
632   &21:25:10.3  &+47:56:32 &15.35  &0.02  &1.22  &0.01 & 0.07  &0.03  &0.57  &0.01  &2.62  &0.00 &0 &J212510.28+475631.1 &21251026+4756311\\

\end{longtable}
\end{landscape}

% \newpage
\begin{longtable}{ccccccc}
\caption{List of B stars selected as members and used to compute the mean absorption of the cluster (see Fig.\,\ref{fig:Vc1B}). 
Only ten rows are presented here. Full table is available as online material.}\\\label{tab:Bmemb}
ID& RA&      DEC&    $A_V$ & $V_0$ &  $c_0$& $(b-y)_0$\\\hline
  648 &  21:25:00.3&+48:00:26&3.03&9.33 &0.08 &-0.10\\
  687 &  21:24:42.0&+47:57:29&3.47&12.42&0.26 &0.02 \\
  696 &  21:24:38.0&+48:00:39&3.55&8.72 &0.12 &0.03 \\
  705 &  21:24:33.5&+48:00:15&2.63&12.59&0.35 &-0.16\\
  706 &  21:24:33.2&+48:00:39&2.72&13.2 &0.65 &-0.12\\
  735 &  21:24:23.4&+48:00:47&2.43&13.49&0.56 &-0.19\\
  741 &  21:24:22.0&+48:00:51&2.35&13.45&0.76 &-0.19\\
  745 &  21:24:20.9&+48:00:33&2.48&11.63&0.37 &-0.19\\
  748 &  21:24:20.3&+48:04:40&2.69&12.41&0.32 &-0.15\\
  755 &  21:24:17.5&+48:01:33&2.45&12.63&0.43 &-0.19\\
\end{longtable}

% \clearpage 
% \onecolumn
\begin{longtable}{ccccccccccc}
\caption{Individual physical parameters computed for stars with $T_{\unit{eff}}>7000\,\unit{K}$ following \citet{2014A&A...568A.119M}. Only ten rows are presented here. Full table is available online.}\\\label{tab:TablePF}
\centering
ID & $T_{\unit{eff}}$ & $\sigma_{T_{\unit{eff}}}$ & $\log g$ & $\sigma_{\log g}$ & Dist & $\sigma_{\unit{Dist}}$&$M_V$&$\sigma_{M_V}$&$A_V$&$\sigma_{A_V}$ \\\hline
 625 &8870    &57   &4.04   &0.03   & 934   &  41   &1.22   &0.08   &2.44   &0.05\\
 628 &7270    &42   &3.43   &0.05   &2100   & 123   &0.39   &0.11   &2.67   &0.04\\
 629 &9050    &39   &4.20   &0.04   &1398   &  69   &1.61   &0.09   &2.24   &0.03\\
 630 &7860    &20   &4.23   &0.02   & 734   &  33   &2.42   &0.05   &2.68   &0.06\\
 633 &7410    &21   &4.02   &0.08   &1969   & 240   &2.12   &0.25   &2.83   &0.06\\
 634 &8350    &37   &4.25   &0.02   & 934   &  25   &2.15   &0.05   &2.25   &0.03\\
 636 &7010    &36   &4.35   &0.04   & 628   &   7   &3.28   &0.03   &2.48   &0.03\\
 639 &7900    &50   &4.29   &0.05   &1923   & 104   &2.56   &0.09   &2.59   &0.05\\
 640 &9160    &35   &4.30   &0.02   &1225   &  34   &1.86   &0.05   &2.26   &0.03\\
 643 &7690    &33   &4.31   &0.07   &1107   &  60   &2.75   &0.10   &3.28  &0.04\\

\end{longtable}

%%%%%%%%%%%%%%%%%%%%%%%%%%%%%%%%%%%%%%%%%%%%%%%%%%

% Don't change these lines
\bsp	% typesetting comment
\label{lastpage}
\end{document}